\documentstyle[aps,prl,twocolumn,epsf]{revtex}

\def\be{\begin{eqnarray}}
\def\ee{\end{eqnarray}}

\begin{document}
\draft

\title{
Thermodynamic origin of universal fluctuations and two-power laws 
}

\author{Jan Naudts$^1$ and Marek Czachor$^{1,2,3}$}
\address{
$^1$ Departement Natuurkunde, Universiteit Antwerpen UIA,
Universiteitsplein 1, B2610 Antwerpen, Belgium\\
$^2$ Katedra Fizyki Teoretycznej i Metod Matematycznych,
Politechnika Gda\'{n}ska, 80-952 Gda\'{n}sk, Poland\\
$^3$ Department of Physics, Technische Universit\"at Clausthal,
38678 Clausthal-Zellerfeld, Germany\\
E-mail: mczachor@pg.gda.pl and Jan.Naudts@ua.ac.be
}

\maketitle

\begin{abstract}

We discuss universality of response functions in systems with 
excited degrees of freedom. We propose a unification of two 
existing phenomenologies, two-power law decay and deviation from 
power law due to non-extensivity. A universal curve is derived 
by maximizing entropy with a non-linear constraint. The same 
formalism can explain the universal 
fluctuation curves which have been discovered recently by 
Bramwell, Holdsworth, and Pinton.

\end{abstract}



Many experiments measure the probability $p(t){\rm d}t$
that an event takes place within the time interval
$[t,t+{\rm d}t]$ after excitation of the system at $t=0$.
Classical arguments predict exponential decay.
We are interested here in a power law decay at large times
\be
p(t)\sim t^{-1/(1-\alpha)}
\label{pldecay}
\ee
with $0<\alpha<1$.

There is an extensive literature explaining non-exponential decay 
by means of fractal theory. In this domain of research it is accepted
that a scaling law like (\ref{pldecay}) holds only
for a limited range $t_1<t<t_2$ of the argument $t$.
For $t<t_1$ collective effects presumably restore non-fractal
behavior. For large $t$ individual behavior
on microscopic scale dominates and non-universal decay
is expected. A nearly ideal example of this behavior has been observed
recently in quantum dots \cite {KFH00}, with a power law extending
over a range of $10^5$ of time scales.
Many theoretical models explain power law decay as the
correct asymptotics in the large time limit.

There is more and more evidence that also non-power law decay
can be described by a universal probability density function
(PDF).
An early explanation \cite {ABE74} of non-exponential decay
involves the notion
of an activation energy required to trigger the decay process.
These authors proposed the following modification
of (\ref{pldecay}) (using our notations)
\be
p(t)&\sim& (1+a(1-\alpha)\omega t)^{-1/(1-\alpha)}
\label{ffdecay}
\ee
with $a$ some positive constant, and $\omega$ a constant with
dimension of a frequency, introduced to
obtain the dimensionless combination $\omega t$.
Later on, (\ref{ffdecay}) lost importance in favor of stretched
exponential decay \cite {KR47}
\be
p(t)\sim e^{-\kappa (\omega t)^\gamma},
\ee
which is expected in case of processes with a scaling distribution
of relaxation rates. Here, we restrict ourselves to the case of
asymptotic decay with a single power law.

In dielectric relaxation one observes usually 
two regions of power law response (in our notations)
\be
p(t)&\sim& (\omega t)^{-1/(1-\alpha)} \qquad \omega t \ll 1\cr
    &\sim& (\omega t)^{-1/(1-\rho)} \qquad \omega t \gg 1.
\label{twodecay}
\ee
(see e.g.~Weron and Jurlewicz \cite {WJ93,JW99}).
The point of view of the present paper is that both behaviors,
as described by (\ref{ffdecay}) and (\ref{twodecay}),
should be combined. More precisely, the generic behavior
in the region $\omega t \ll 1$ should be described by (\ref{ffdecay})
instead of (\ref{pldecay}). This substitution makes sense
because (\ref{ffdecay}) describes a curve which starts off
linearly and bends over towards a power law decay with exponent
$-1/(1-\alpha)$.

\begin{figure}
\epsfxsize=8.25cm
\epsffile{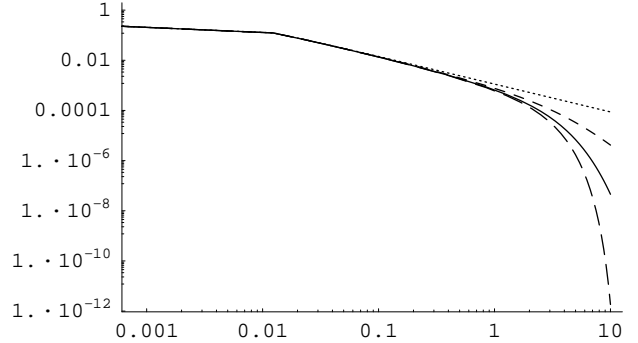}
\caption{Log-log-plot of $p(t)$ with $a=500$ for different
combinations of exponents $(\alpha,\rho)$:
(0.1,0.1) dotted, (0.1,1) solid, (0.1,0.8) short-dashed,
(0.1,1.08) long-dashed.}
\end{figure}

The actual formula, which we will derive below,
and which covers the whole fractal region, is (see Fig. 1)
\be
p(t)\sim
\left(1-a+a\left(
1+(1-\rho)\omega t
\right)^\frac{1-\alpha}{1-\rho}
\right)^{-1/(1-\alpha)}.
\label{pdf}
\ee
Three regions can be distinguished.
In regions 1 and 2 the decay is as described by (\ref{ffdecay}).
Regions 2 and 3 show power law decay according to (\ref{twodecay}).
Note that $\rho>1$ corresponds to a super-exponential decay.

For $\rho=\alpha$ expression (\ref{pdf}) reduces to (\ref{ffdecay}).
If $a=1$ then it reduces to (\ref{ffdecay}) with $\alpha$
replaced by $\rho$.
In the limit $\rho=1$ one obtains
\be
p(t)\sim
\left(1-a+a\exp((1-\alpha)\omega t)
\right)^{-1/(1-\alpha)}.
\label {rendecay}
\ee
In the limit $\alpha=1$ one obtains
\be
p(t)\sim
(1+(1-\rho)\omega t)^{-a/(1-\rho)}.
\ee
In any case, in the limit $\alpha=\rho=1$
the decay is exponential $p(t)\sim \exp(-a\omega t)$.

We found first evidence for (\ref{pdf}) in the work
of Tsallis, Bemski, and Mendes \cite{TBM99}. They reanalyze
the experimental data of \cite {ABE74} and
observe that the description using
(\ref{ffdecay}) improves significantly by using
(\ref{rendecay}), formula which was introduced on an {\sl ad hoc}
basis.  The full data set of \cite {ABE74} could be described
with it. The universality of (\ref{rendecay}) is further supported
by recent work of Montemurro \cite {MM01}
using a very convincing analysis of linguistic data.
In both papers \cite{TBM99} and \cite {MM01} there is evidence
that the asymptotic behavior for large $t$ still deviates from
(\ref{rendecay}), and follows a power law instead of decaying exponentially.
In the logic of the present paper, this is a cross-over from
one power law to another, as described by (\ref{twodecay}).
Hence (\ref{pdf}) should be used with $\rho\not=1$
instead of (\ref{rendecay}). In \cite {TBM99} an {\sl ad hoc} modification of
(\ref{rendecay}) was proposed, slightly different from (\ref{pdf}).
The same expression was used in \cite {MM01} and fits
the experimental data over the whole range of $t$-values.
It is clear that (\ref{pdf}) will fit as well, but has a number of
advantages:
1) there is a general formalism, which is used to derive it; 
2) the PDF has a workable algebraic
expression, which is not the case for the formula
proposed in \cite {TBM99};
3) within the context of the formalism an average time
$\langle\langle \omega t\rangle\rangle$ is defined 
using a nonlinear average -- see (\ref{knav}) below.
The average time characterizes the PDF $p(t)$ together with the 
two exponents $\alpha$ and $\rho$, and the cross-over frequency 
$\omega$. It replaces the fitting parameter $a$ which has a less 
clear physical meaning (see Fig.~2).
In particular, in the case $\rho=1$, we 
show \cite {NC01} that $\langle\langle\cdot\rangle\rangle$ satisfies
a property of additivity, which implies that it is meaningful 
to add average times of different experiments all having the 
same values of $\alpha$ and $\omega$.

\begin{figure}
\epsfxsize=8.25cm
\epsffile{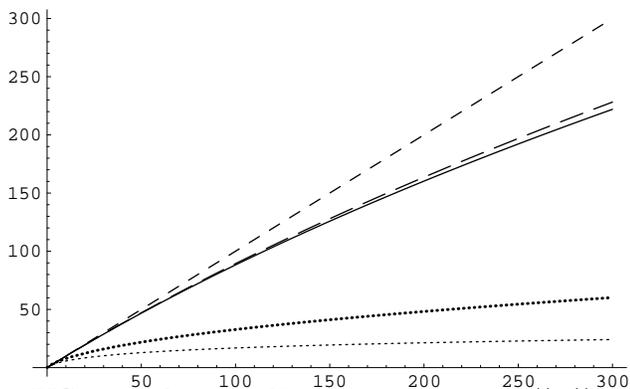}
\caption{Kolmogorov-Nagumo average time
$\langle\langle \omega t\rangle\rangle$ as a function of $1/a$
for several values of $(\alpha,\rho)$:
$(0.9999,0.9999)$ straight line, short-dashed;
$(0.999,0.9999)$ long-dashed;
$(0.999,1)$ solid line;
$(0.8,0.9)$ thick-dotted;
$(0.9,0.9999)$ thin-dotted. 
}
\end{figure}

Up to now we considered decay as a function of time. However the 
PDF (\ref{pdf}) can also be considered as a generalization of 
the distribution of Boltzmann-Gibbs. Time $t$ is then 
replaced by energy $E$, frequency $\omega$ by a fixed inverse 
temperature $\beta$. With this interpretation (\ref{ffdecay}) is 
the equilibrium distribution of non-extensive thermostatistics 
\cite {TC88}. The entropic parameter $q$ of this formalism,
coincides with the exponent $\alpha$ (the relation becomes
$q=2-\alpha$ in the recently
modified formalism \cite {TMP98}).
In the context of non-extensive thermostatistics 
(\ref{ffdecay}) has numerous applications
--- for a review see \cite {TC99} or \cite {AO00}. 
In this context 
(\ref{rendecay}) becomes the equilibrium distribution
for thermostatistics based on R\'enyi's choice of Komogorov-Nagumo
averages and $\alpha$-entropies -- see \cite {NC01}. 


Let us now discuss universal distributions of fluctuating
quantities. Their universality has been discovered
recently by Bramwell, Holdsworth, and Pinton (BHP) \cite {BHP98,PH99},
based on earlier work \cite{LPF96}. 
The result received ample support in the literature
\cite{HKG99,B00,AG01,BFH01,CRW01,DJ01,DM01,PHSB01,ADG01}.
The BHP density function is of the form 
\be
g_{_{\rm BHP}}(\epsilon)\sim \exp\left(b(y-e^y\right)
\label {univspec}
\ee
with $y=c(\epsilon-u)$.
Setting $b=1$
one finds the famous Fisher-Tippett density
appearing in statistics of extremes (see Gumbel \cite {GEJ58}).
However, the value which empirically best describes
fluctuation spectra is $b=\pi/2$.

Using the standard central limit theorem (CLT) we show at the end of this
Letter that our formalism implies (for $\rho>2/3$) 
\be
{}&{}&
\tilde g(\epsilon)=\Big(\exp_\rho\big[v
+\epsilon\big]\Big)^{\rho-\alpha}\nonumber\\
&{}&\phantom{===}\times
g\Big(\frac{\sqrt{N}}{\sigma}\big[\phi_{\alpha\rho}(v+\epsilon)-
\phi_{\alpha\rho}(v)
\big]\Big).\label{flucspec}
\ee
with $\sigma^2$ the variance of $p(t)$,
$g(x)$ the normal density, and the functions 
$\phi_{\alpha\rho}$ and $\exp_\rho$ defined later in the text
(for $\rho<2/3$ 
the normal density function $g(\epsilon)$ should be replaced by a
L\'evy-stable PDF). 

Of particular interest is the limit $\rho=1$ (corresponding to R\'enyi
$\alpha$-entropies -- see further on). 
Then (\ref{flucspec}) can be written as
\be
\tilde g(\epsilon)\sim\exp\left(y-\frac{1}{2}d^2(e^y-1)^2\right)
\label{fluctext}
\ee
with $y=(1-\alpha)\epsilon$ and
$d=\exp((1-\alpha)v)/N^{1/2}\sigma(1-\alpha)$.
This expression differs in an essential way from (\ref{univspec}).
However, both PDFs have a similar shape (see Fig.~3).
Note that (\ref{univspec}) has an additional
fitting parameter, the role of which is taken over here
by the exponent $\rho$. Very accurate
data will be needed to distinguish the two PDFs
on an experimental basis.
An advantage of (\ref{fluctext}) is that in the limit
$\alpha=1$ it reduces to the normal density with width
$\sigma$, as expected.
Note also that, if $g(\epsilon)$ were replaced by the exponential density,
and $\rho=1$, then (\ref {flucspec}) reduces to (\ref {univspec})
with $b=1$ and $c=1-\alpha$.
However, the exponential density is {\sl not}
L\'evy-stable. Therefore we do not expect it on thermodynamic grounds.

\begin{figure}
\epsfxsize=8.25cm
\epsffile{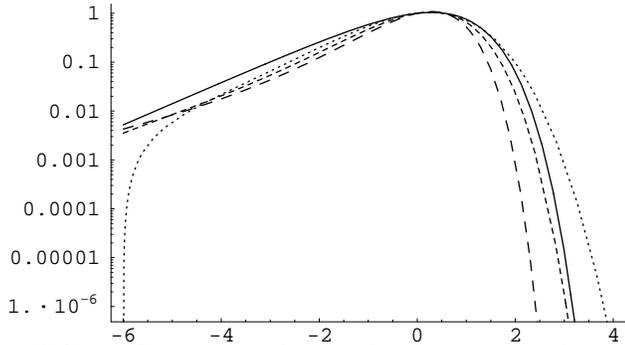}
\caption{
Comparison of $\tilde g(\epsilon)$ 
and $g_{_{\rm BHP}}(\epsilon)$:
$\tilde g(\epsilon)$ with $\rho=0.8$ dotted,
with $\rho=0.9$ short-dashed,
with $\rho=1$ long-dashed;
$g_{_{\rm BHP}}(\epsilon)$ solid;
the curves are normalized to 1 in the origin;
parameters of $\tilde g(\epsilon)$  are $v=1$, $\alpha=0.5$, and
$\sigma^2=4N$;
the parameters of $g_{_{\rm BHP}}(\epsilon)$ have been adapted to fit roughly
$\tilde g(\epsilon)$ with $\rho=1$.
}
\end{figure}


In the final part of this Letter we derive formulas (\ref{pdf})
and (\ref{flucspec}) using thermodynamic arguments.
Weron and Jur\-le\-wicz \cite {WJ93,JW99} express the response function $p(t)$ 
in terms of PDFs $p_i(t)$, describing relaxation
of individual dipoles. To do this, they use statistics of extremes \cite {GEJ58}.
The underlying assumption is that the whole system relaxes as soon
as one of the individual dipoles relaxes. Our assumptions are not
so drastic. We start with a thermodynamic argument,
and replace statistics of extremes by the usual CLT.

About 20 years ago Montroll and Shlesinger \cite {MS83}
proposed a thermodynamic derivation of Einstein's
diffusion law. It follows by optimizing entropy of the PDF $p(t)$
under the constraint of zero first moment and fixed second moment.
They observed that anomalous diffusion can be explained
along similar lines using an {\sl ad hoc} constraint
replacing the constraint on the second moment.
Alemany and Zanette \cite {AZ94,ZA95},
followed by Tsallis et al \cite {TLS95,PT99},
repeat these arguments,
replacing Shannon's entropy by the generalized entropy
used in nonextensive thermodynamics \cite {TC88}
\be
S_\alpha(p)=\int\hbox{ d}t\,\frac{p(t)^\alpha-p(t)}{1-\alpha},
\qquad \alpha>0,\alpha\not=1.
\label {tsallis}
\ee
These authors introduce a suitably adapted constraint on the second moment.
Recently \cite {NC01}, the present authors generalized
nonextensive thermostatistics by founding it on nonlinear
Kolmogorov-Nagumo averages \cite{KA30,NM30}
\be
\langle\langle f\rangle\rangle
=\phi^{-1}\left(\int\hbox {d}t\,p(t)\phi(f(t))\right).
\label{knav}
\ee
This average depends on a monotonically increasing function $\phi(x)$.
In the present paper we choose $\phi=\phi_{\alpha\rho}$ with
\be
\phi_{\alpha\rho}(x)&=&\ln_\alpha(\exp_\rho(x))\cr
&=&\frac{1}{1-\alpha}\left[
(1+(1-\rho)x)^{(1-\alpha)/(1-\rho)}-1
\right].
\ee
The $\alpha$-deformed exponential and logarithmic functions are
defined by \cite {TC94,BEP98}
\be
\exp_\alpha (x)&=&\left[1+(1-\alpha)x\right]^{1/(1-\alpha)}\cr
\ln_\alpha (x)&=&\frac{x^{1-\alpha}-1}{1-\alpha}.
\ee
The inverse function is $\phi_{\rho\alpha}(x)$.
Following \cite {NC01}, the corresponding definition of entropy is
\be
S_{\alpha\rho}(p)=\phi_{\rho\alpha}\left(\int\hbox{ d}t\,p(t)\ln_\alpha(1/p(t))
\right).
\ee
In the limit $\rho=1$ this is the $\alpha$-entropy of R\'enyi \cite {RA60}.
\be
S_{\alpha 1}(p)=\frac{1}{1-\alpha}\ln\left(
\int\hbox{ d}t\,p(t)^\alpha
\right).
\ee
On the other hand, one has $\phi_{\alpha\alpha}(x)=x$ so that
$S_{\alpha\alpha}(p)$ coincides with (\ref{tsallis}).
Let us now optimize entropy $S_{\alpha\rho}(p)$ under the constraint
that the average decay time $\langle\langle \omega t\rangle\rangle$
has a given value. A straightforward calculation using
Lagrange parameters produces an implicit expression
for the PDF $p(t)$. By introducing a free parameter $a$
the latter can be made explicit, resulting in (\ref{pdf})
(see \cite{NC01}).

In order to calculate the fluctuation spectrum assume that $\rho>2/3$ 
(then the second moment of $p(t)$ exists and CLT
holds). If $A(t)$ is an arbitrary random variable we will denote its {\it
linear\/} (experimental and theoretical) averages by
\be
\langle A\rangle_N
=
\sum_{k=1}^N A(t_k)/N,\quad
\langle A\rangle
=
\int \hbox{ d}t\, p(t)A(t)\nonumber
\ee
Denote by $P(x)$ the probability of an event $x$,
given the PDF $p(t)$. In this notation CLT can
be formulated as
\be
P\Big(\langle A\rangle_N\leq z\Big)
&\approx&
F\Big(\frac{\sqrt{N}}{\sigma}(z-\langle A\rangle\Big)
\ee
with $\sigma^2$ the variance of $p(t)$,
$F$ the normal distribution function,
and $N$ sufficiently large \cite{Feller}. 

The nonlinear
averages 
\be
\langle\langle A\rangle\rangle_N
=
\phi^{-1}
\Big(\langle \phi(A)\rangle_N\Big),\quad
\langle\langle A\rangle\rangle
=
\phi^{-1}
\Big(\langle \phi(A)\rangle\Big)\nonumber
\ee
are related via $\phi^{-1}$ 
to the linear ones of the random variable
$B(t)=\phi\big(A(t)\big)$. Since $\phi$ is monotonically increasing we can
write 
\be
P\Big(\langle\langle A\rangle\rangle_N\leq \langle\langle A\rangle\rangle
+\epsilon\Big)
&=&
P\Big(\langle \phi(A)\rangle_N\leq \phi\big(\langle\langle A\rangle\rangle
+\epsilon\big)\Big)\nonumber
\ee
and apply CLT to its right side. One finds
\be
{}&{}&
P\Big(\langle\langle A\rangle\rangle_N\leq \langle\langle
A\rangle\rangle
+\epsilon\Big)\nonumber\\
&{}&\phantom{==}
\approx
F\Big(\frac{\sqrt{N}}{\sigma}\big[\phi\big(\langle\langle A\rangle\rangle
+\epsilon\big)-
\phi\big(\langle\langle A\rangle\rangle\big)\Big).
\ee
Introduce a limiting PDF $\tilde g(\epsilon)$ by 
\be
\frac{d}{d\epsilon}
P\Big(\langle\langle A\rangle\rangle_N\leq
\langle\langle A\rangle\rangle +\epsilon\Big)
&\approx&
\frac{\sqrt{N}}{\sigma}\tilde g(\epsilon).
\ee
Using $g(x)=dF(x)/dx$ and choosing $\phi=\phi_{\alpha\rho}$ we finally
obtain
\be
{}&{}&
\tilde
g(\epsilon)=\Big(\exp_\rho\big[\langle\langle
A\rangle\rangle
+\epsilon\big]\Big)^{\rho-\alpha}
\nonumber\\
&{}&\phantom{===}\times
g\Big(\frac{\sqrt{N}}{\sigma}\big[\phi\big(\langle\langle A\rangle\rangle
+\epsilon\big)-
\phi\big(\langle\langle A\rangle\rangle\big)
\big]\Big).
\ee
It is clear that for $\alpha=\rho$ one gets
$\tilde g(\epsilon)=g(\sqrt{N}\epsilon/\sigma)$. 
Setting
$\langle\langle A\rangle\rangle=v$ we obtain
(\ref{flucspec}).  


Let us summarize the results. Applying the standard formalism
of maximum entropy
principles but formulated in terms of nonlinear averages we can explain a
large class of
response functions with asymptotic power law decay, including the 
ubiquitous two-power law. Moreover, using the same formalism  
and standard central limit theorem we obtain a 
universal behavior of fluctuation spectra. In both cases the universal
laws we propose differ slightly from the known ones. To
distinguish experimentally between them one needs more
precise experiments.


The stay of MC at Universiteit Antwerpen, UIA, was made possible 
by NATO through a research fellowship. Part of the results were 
obtained during our visit at Technical University of Clausthal. 
We are indebted to Prof. H. D. Doebner for his comments and his 
interest in this work. MC thanks the Alexander von Humboldt 
Foundation for support. 


\end{document}